\documentclass[oldversion]{aa}
\usepackage{graphicx}
\begin{document}
   \title{Integral field spectroscopy of L 449-1
\thanks{Based on observations collected at the European Southern Observatory, 
Chile (ESO No.\ 078.C-0540).}
}

   \subtitle{A test case for spectral differential imaging with SINFONI}

   \author{Markus Janson\inst{1} \and
          Wolfgang Brandner\inst{1} \and
	  Thomas Henning\inst{1} 
          }

   \offprints{Markus Janson}

   \institute{Max-Planck-Institut f\"ur Astronomie, K\"onigstuhl 17,
              D-69117 Heidelberg, Germany\\
              \email{janson@mpia.de, brandner@mpia.de, henning@mpia.de}
             }

   \date{Received ---; accepted ---}

   \abstract{
Spectral differential imaging is an increasingly used technique for ground-based direct imaging searches for brown dwarf and planetary mass companions to stars. The technique takes advantage of absorption features that exist in these cool objects, but not in stars, and is normally implemented through simultaneous narrow-band imagers in 2 to 4 adjacent channels. However, by instead using an integral field unit, different spectral features could be used depending on the actual spectrum, potentially leading to greater flexibility and stronger detection limits. In this paper, we present the results of a test of spectral differential imaging using the SINFONI integral field unit at the VLT to study the nearby active star L449-1. No convincing companion candidates are found. We find that the method provides a $3 \sigma$ contrast limit of 7.5 mag at 0.35", which is about 1.5 mag lower than for NACO-SDI at the same telescope, using the same integration time. We discuss the reasons for this, and the implications. In addition, we use the SINFONI data to constrain the spectral type in the NIR for L 449-1, and find a result between M3.0 and M4.0, in close agreement with a previous classification in the visual range.
   
   }{}{}{}{}

   \maketitle
%

\section{Introduction}

In recent years, developments in techniques and instrumentation have led to a strongly increased capacity for high-contrast imaging of substellar companions at small angular separations to stars, from the ground. In particular, the use of high-order adaptive optics (AO) in combination with various differential imaging techniques provides a sensitivity for companions that are $10^4$-$10^5$ times fainter than the primary at a separation of 0.5"-1" at near-infrared wavelengths (see Janson et al. 2007 and Biller et al. 2007). According to theoretical models (see e.g. Baraffe et al. 2003), this corresponds to objects of a few times the mass of Jupiter around young stars. 

A particularly efficient technique for such observations is simultaneous spectral differential imaging (SDI, see e.g. Rosenthal et al. 1996 and Racine et al. 1999), in which the flux from a star is observed in two narrow, adjacent wavelength bands simultaneously. The bands are chosen such that one is inside and one outside of an absorption feature which arises uniquely in cool atmospheres. One example of such a feature is the methane band at 1.6 $\mu$m, which only occurs in dwarfs of spectral type T or later, whereas a stellar spectrum is flat in this range. By subtracting one image from the other, the stellar point spread function (PSF) can be largely subtracted out, since it is approximately flat around 1.6 $\mu$m. Along with the main PSF, most of the random PSF substructure (speckle noise) is subtracted out as well, since the images are simultaneous and hence the atmospheric speckle pattern is the same in both images.

NACO-SDI at the VLT is an excellent instrument for the purpose of SDI observation (see Lenzen et al. 2004). NACO-SDI images the same field in three different narrow bands around the 1.6 $\mu$m methane feature simultaneously. It is designed to minimize non-common path aberrations, and has been shown to be capable of achieving a contrast of more than 13 mag between star and companion at 1" angular separation at the 3$\sigma$ level (see Janson et al. 2007), without the use of a coronagraph. One limitation of the NACO-SDI instrument is that it can only attain information about the methane feature in question. It has been suggested (Berton et al. 2006) that if $N > 1$ different absorption features are used at once for SDI purposes, the achievable contrast would increase by a factor $N^{1/2}$ for the same integration time, simply due to the more efficient use of photons. Hence, in principle, an instrument with the capacity to do multi-feature SDI (MSDI) would be preferable to NACO-SDI, if the instruments perform equally well in all other respects (in terms of differential aberrations, etc.).

SINFONI is an integral field spectroscopy instrument with AO capability at the VLT (see Eisenhauer et al. 2003 and Bonnet et al. 2004). It uses an image slicer to divide the FOV into pseudo-slits that are individually dispersed on a grating, hence retaining both the spatial and spectral information of the incoming light. The output data can be used to construct a data cube that stores spatial information along two axes, and spectral information along the third. In this way, SINFONI can be used as an SDI or MSDI instrument, since the cube contains simultaneous narrow-band images over a large range of wavelengths.

We have acquired SINFONI data to test its capacity for SDI. In the following sections, we describe the observations and data reduction, and compare the results with NACO-SDI. We also compare the results to a spectral deconvolution (SD) scheme applied to another set of SINFONI data by Thatte et al. (2007). We dicuss the advantages and disadvantages of using SINFONI for high-contrast imaging and characterization of exoplanet and brown dwarf companions to stars. Finally, we analyze the non-differential collapsed images, and constrain the spectral type of L 449-1 from H- and K-band spectroscopy.

\section{Observations}

The target chosen for the observation was L 449-1. This star was identified as a high proper motion star in Scholz et al. (2005), who also used spectroscopy and photometry to classify the object as an M4 star only 5.7 pc away. In addition, the star shows signs of activity, through both H$\alpha$ emission and identification with a bright X-ray source, possibly indicating a young age. Hence, it is of particular interest for high-contrast imaging surveys, but since it was not properly classified prior to the publication of Scholz et al. (2005), it has not been targeted by such surveys. For this reason, L 449-1 makes an excellent test case for high-contrast imaging with SINFONI.

The data of L 449-1 were taken using SINFONI at the VLT during a visitor mode run at Paranal on the night of 26 Jan 2007. In order to get the best possible spatial sampling, the smallest available SINFONI field of view of $0.8" \times 0.8"$ was used, yielding an effective pixel scale of 12.5 mas/pixel by 25 mas/pixel. The ``H+K'' setting was used, covering the spectral range of 1.4 $\mu$m to 2.5 $\mu$m at a spectral resolution of about $R=1500$. Adaptive optics was used with the target itself as the guide star, and provided good and stable wavefront correction -- the average Strehl ratio was 34 \% (corresponding to a wavelength of 1.95 $\mu$m) as given by the AO system itself. The weather conditions were also good and stable, with an average seeing of 0.72" at 500 nm as given by the atmospheric seeing monitor at Paranal.

For the purpose of efficient subtraction of the stellar PSF, the data were taken at two different instrument rotations. This enables use of a data reduction scheme known as roll subtraction (see Mueller \& Weigelt 1987) or angular differential imaging (ADI, see e.g. Marois et al. 2006). Our implementation of this technique is described in Kellner (2005) and Janson et al. (2007). Five target frames and one sky frame were taken at an angle of $0^{\circ}$, and three target frames plus one sky frame at an angle of $90^{\circ}$. The effective integration time per frame was 350 seconds.


\section{Data reduction}

All the frames generated during observations were translated into equivalent data cubes, with spatial information along two of the axes, and spectral information along the third, using the standard ESO data reduction pipeline with the Gasgano GUI tool. This procedure also performed all basic data reduction steps, such as sky subtraction, flat fielding, bad pixel correction and wavelength calibration. The output data cubes were used in a number of further data reduction procedures to generate useful data products, using IDL routines that were written specifically for these purposes. The results of the basic reduction also included extracted spectra of L449-1 and the telluric standard star, that were also further processed by our IDL routines for spectral type analysis. We note that this procedure, which is detailed below, was designed to correspond as closely as possible to the equivalent procedure for NACO-SDI (see Kellner 2005 and Janson et al. 2007) so that no difference in the final performance can be simply attributed to a difference in the implementation of the technique.

For the SDI, which was our primary purpose, different combinations of image slices were co-added to represent images of different filters. In particular, images denoted by $f_1$, $f_2$ and $f_3$ were created, where $f_1$ is an image in the wavelength range 1.5625 $\mu$m to 1.5875 $\mu$m, $f_2$ in 1.5875 $\mu$m to 1.6125 $\mu$m, and $f_3$ in 1.6125 $\mu$m to 1.6375 $\mu$m. This construct was designed in order to create, as close as possible, equivalence to the $F_1$, $F_2$ and $F_3$ filters of NACO-SDI (see Janson et al. 2007). The PSF center was determined for every image slice in every data cube through gaussian centroiding, and the images were shifted to a common position based on this information. Remaining differences in photocenter positions were minimized via cross-correlation. Sub-pixel shifts were accommodated through bilinear interpolation. The data were also subsampled by a factor 2 to get the same effective sampling in the x- and y-direction, and to avoid image artefacts arising from the double representation of each sample point in the y-direction (each image slice is 64 by 64 pixels in a SINFONI data cube, but since the actual sampling is 12.5 mas/pixel in the x-direction and 25.0 mas/pixel in the y-direction, each sample point is represented twice in the y-direction). $f_1$, $f_2$ and $f_3$ were subsequently rescaled to a common $\lambda / D$ scale and subtracted from each other to produce SDI images. Unsharp-masking was applied in order to remove the low spatial frequencies (background and halo), whilst retaining higher-frequency features (such as possible companions, but also speckles). This was done by convolving each image with a gaussian kernel with a FWHM of 140 mas, and subsequently subtracting the smooth image from the original image.

Finally, the resulting difference images at each rotation angle were co-added, after which, any co-added image corresponding to one angle was subtracted from the other. The radial profile of the error was calculated for the final image. This was done for each separation by selecting all pixels between an inner and outer radius of $\pm 1$ pixel from the separation of interest, and taking the standard deviation of those pixels. By comparing this radial profile to the brightness of the primary star itself, the achieved contrast as a function of angular separation could be determined.

For test purposes, the same procedure as described above was also performed for three images in the K-band: $g_1$ at 2.095 $\mu$m to 2.120 $\mu$m, $g_2$ at 2.120 $\mu$m to 2.145 $\mu$m and $g_3$ at 2.145 $\mu$m to 2.170 $\mu$m (i.e., with the same type of configuration and the same bandwidth as $f_1$, $f_2$ and $f_3$). These wavelengths are arbitrary from a physical viewpoint, and are not expected to correspond to any particularly interesting spectral features. Instead, the purpose of the analysis, in this case, was to test whether the contrast performance is significantly different in the H- and K-bands. One particular reason for this was to test whether the output quality is strongly dependent on the spatial sampling (which is sub-critical in the y-direction in the H-band, but critical in the K-band).

For general (non-differential) analysis of the primary itself, as well as for detecting possible relatively low-contrast companions and measuring their spectra, it is most practical to keep all the image slices in the cube at their original $\lambda / D$ scale, and simply do the spatial and spectral analysis based on the original cube. For this purpose, we generated wavelength-collapsed frames in the H- and K-bands in order to produce broadband-equivalent images for spatial analysis. Furthermore, in both H- and K-bands a spectrum integrated over the whole field of view (except for a margin of four pixels at the edges of the field) was calculated. If L 449-1 is single or significantly brighter than any of its companions, the extracted spectrum corresponds to the spectrum of L 449-1 itself. In addition, a new cube was generated by subtracting the extracted spectrum from each spatial position in the original cube, after normalizing the spectrum by the intensity of the broad-band frame at each corresponding position. The result is essentially a differential spectroscopy cube, mapping the spatial distribution of spectral deviance from the mean spectrum.

\section{Results and discussion}

\subsection{Spectral differential imaging}

\begin{figure*}[htb]
   \centering
   \includegraphics[width=17cm]{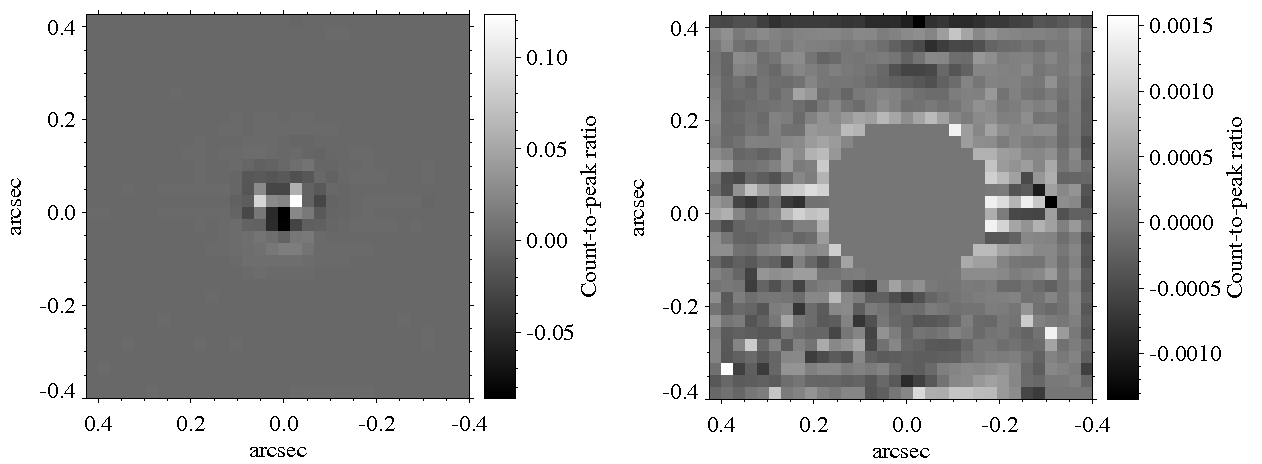}
\caption{SDI image given by $f_1 - f_3$. Left: Low-contrast version to show the central part of the image. Right: High-contrast version with an artificial mask placed over the center, to clearly illuminate the outer part. No strong cool companion candidates can be found in the image. North is up and East is to the left.}
\label{sdi_img}
    \end{figure*}

\begin{figure}[htb]
   \centering
   \includegraphics[width=8.5cm]{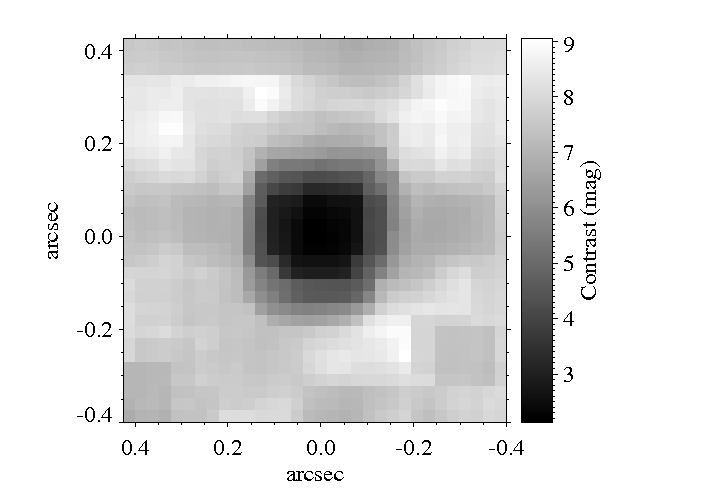}
\caption{SDI $3 \sigma$ contrast map corresponding to Fig. \ref{sdi_img}. Each pixel corresponds to three times the standard deviation of a 5 by 5 pixel square centered on the equivalent position in the SDI image. North is up and East is to the left.}
\label{sdi_map}
    \end{figure}

\begin{figure}[htb]
   \centering
   \includegraphics[width=8.5cm]{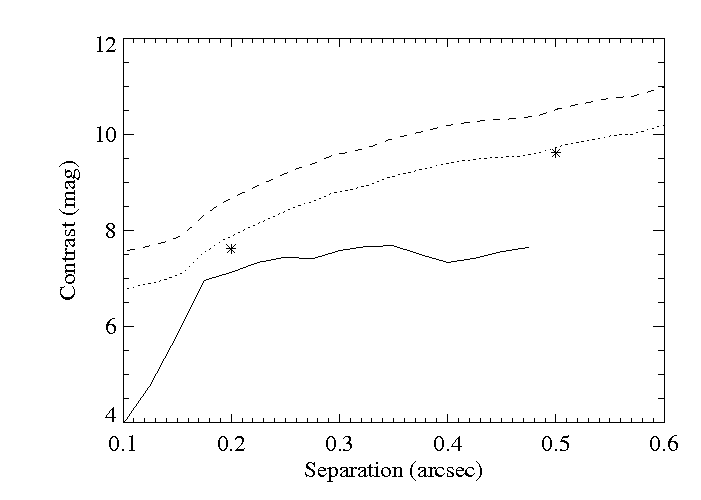}
\caption{Contrast curve for the SINFONI SDI image (solid line) compared to a corresponding NACO-SDI curve for $\epsilon$ Eri (dashed line), and the same curve normalized to the same integration time, assuming a $t^{1/2}$ SNR development (dotted line). Also plotted are two values quoted by Thatte et al. 2007 for spectral deconvolution with SINFONI, renormalized to the same integration time. See the text for discussions. See also Fig. \ref{sdi_phys_err} for a physical example.}
\label{sdi_err}
    \end{figure}

\begin{figure}[htb]
   \centering
   \includegraphics[width=8.5cm]{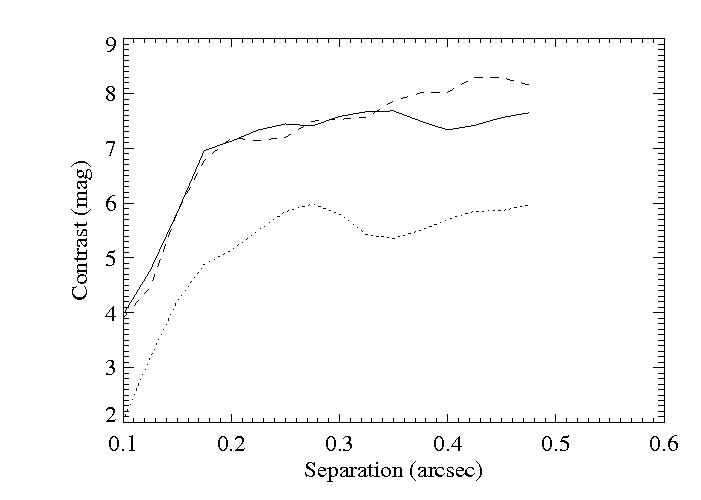}
\caption{$3 \sigma$ Contrast curve for SINFONI SDI for $f_1 - f_3$ (solid line), and for $g_1 - g_3$ (dashed line). The latter is chosen only to show the instrumental performance in the K-band, and is not expected to correspond to any physically useful feature for SDI. The performance is very similar except at the outer edge. Also plotted is the contrast curve for the non-SDI $f_1$ image (dotted line), illustrating that a substantial improvement is gained from the differential imaging methods that we apply.}
\label{hk_comp}
    \end{figure}

The $f_1 - f_3$ image, which corresponds to the standard final output from NACO-SDI, is shown in Fig. \ref{sdi_img}. No interesting companion candidates can be seen in the image. The corresponding $3 \sigma$ contrast map is shown in Fig. \ref{sdi_map}; the $3 \sigma$ radial contrast curve is shown compared to NACO-SDI in Fig. \ref{sdi_err}, and compared to other implementations of the same data set in Fig. \ref{hk_comp}.  It is quite clear that for the same integration time, and virtually the same filter set, and under similar weather conditions, a SINFONI-based SDI approach performs significantly worse than a NACO SDI-based one. While the Strehl ratio is better for NACO-SDI, the seeing is worse, reflecting a difference in AO performance. However, we note that the difference in Strehl ratio is only a factor 1.4, which, if we assume a linear dependence between $SNR$ and Strehl ratio for the NACO-SDI data (see Janson et al., 2007) corresponds to 0.4 mag in brightness contrast. The difference in Fig. \ref{sdi_err} is much larger than that, hence it is not plausible that it is solely due to AO performance. Furthermore, Fig. \ref{hk_comp} shows that the performance over most of the parameter range is very similar in the H- and K-bands, despite the Strehl ratio being higher in the K-band than in the H-band (and in fact, becomes higher than the Strehl ratio for the NACO-SDI data). This again demonstrates that AO performance is not the limiting factor for the achieved contrast of SINFONI SDI. Aside from the average noise impact, the spatial distribution of the noise is also different, as SINFONI SDI has a spatially constant noise floor that dominates most of the image space.

\begin{table*}[htb]
\caption[]{Comparison of the three data sets discussed in the text.}
         \label{astrometry}
\begin{tabular}{cccccccc}
  Name & Date & Mean Seeing & Mean $t_{\rm c}^a$ & Mean Strehl & $t_{\rm full}^b$ & $t_{\rm eff}^c$ & Correction$^d$ \\ 
       &      &             & [s]                &[1.95 $\mu$m] & [s]            & [s]             &                \\ 
            \noalign{\smallskip}
            \hline
            \noalign{\smallskip}
SINFONI SDI & 2007 Jan 26 & 0.72" & 37.6 & 34 \% & 1050 & 1050 & 1.00 \\
NACO-SDI & 2006 Jan 01 & 0.82" & 12.0 & 48 \% & 4472 & 1050 & 2.06 \\
SINFONI SD & 2006 Jan 25 & 0.64" & -- & 37 \% & 1200 & 1050 & 1.07 \\ 
            \noalign{\smallskip}
\end{tabular}
\begin{list}{}{}
\item[$^{\mathrm{a}}$] Correlation time $t_{\rm c}$ given by the AO system, not measured for the SINFONI SD run.
\item[$^{\mathrm{b}}$] The full integration time of each run.
\item[$^{\mathrm{c}}$] The effective comparative integration time of each run assuming ${\rm SNR} \sim t^{1/2}$.
\item[$^{\mathrm{d}}$] The ${\rm SNR}$ correction factor given by $(t_{\rm full}/t_{\rm eff})^{1/2}$
\end{list}
\end{table*}

\begin{figure}[htb]
   \centering
   \includegraphics[width=8.5cm]{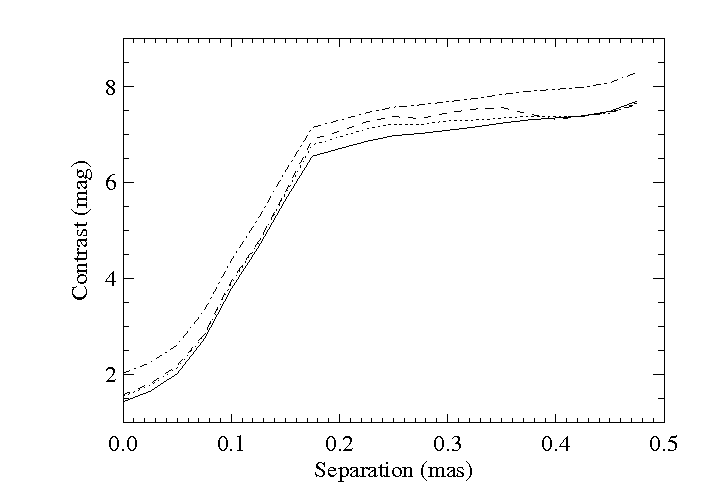}
\caption{Contrast development as a function of time after application of SDI and ADI. Solid line: average curve for one frame per angle. Dotted line: average curve for two frames per angle. Dashed line: average curve for three frames per angle. Dash-dotted line: expectation for the case of three frames per angle, if the noise had been entirely dynamic. It can be clearly seen that the development is considerably slower than in the dynamical case for almost the entire image range, hence quasi-static noise sources are significant.}
\label{err_evol}
    \end{figure}

In Janson et al. (2007), it was shown that the SNR increased with integration time $t$ as almost ${\rm SNR} \sim t^{1/2}$ for NACO-SDI, even for the deepest integrations for most of the spatial range. It is highly interesting to make a similar test for SINFONI SDI, to assess its appropriateness for high-contrast imaging. Such a task is somewhat more challenging for the SINFONI data, since the overheads and bright object constraints of SINFONI forced us to use a poor temporal resolution during the observations, with 5 individual data cubes at $0^{\circ}$ rotation angle and 3 at $90^{\circ}$. To get a robust picture of how the SNR develops, we calculated every possible combination of frames for one, two and three frames per angle, and averaged the contrast curves for each case. The result is shown in Fig. \ref{err_evol}. It is clear that the SNR develops considerably slower than what would be expected if the noise was dominated by temporally uncorrelated sources. This shows that increasing the integration time would not improve the results to the same degree as is the case for NACO-SDI. 

The relatively poor performance and quasi-static nature of the residual noise, as well as its spatial distribution, implies that errors dominate in the SINFONI data that are not present or are well controlled for NACO-SDI. For SINFONI, the image is divided into horizontal stripes by the image slicer, after which, each slice is dispersed on a grating and subsequently registered on the detector. Hence, each spatial slice travels a slightly different optical path, and additionally, can suffer from edge effects in the interaction with the image slicer. Also, each wavelength travels a different optical path after being dispersed on the grating. These differential effects lead to quasi-static errors that may dominate the noise. The horizontal stratification that can be traced in parts of the image certainly implies that the image slicer is at least partly responsible for the final result. Path differences for light of different wavelengths also occur in NACO-SDI for the different sub-images, but in that case, a primary objective was to minimize precisely such errors (see e.g. Brandner et al. 2004). Any possible problems with the cube generation, as peformed by the ESO pipeline, could also affect the results. The similar results achieved in the H- and K-bands (Fig. \ref{hk_comp}) implies that spatial sampling is not a critical issue for this case.

Since the dominating error is (quasi-)static rather than dynamic, adding more spectral features will not increase the achievable contrast to the degree expected in a dynamical case. Aside from this, we note that no other feature apart from the CH$_4$ band in the H- or K-bands is quite as narrow, or has quite as high contrast between the continuum and the full depth of the feature. Therefore, we have to conclude that SDI with SINFONI cannot perform as well as NACO-SDI in terms of detecting previously unknown substellar companions. However, for follow-up observations of high-contrast objects detected by other means, where spectral properties need to be determined, the method has excellent prospects.

On this subject, we will briefly discuss the spectral deconvolution (SD) technique. SD was suggested for detection of substellar companions by Sparks \& Ford (2002), and has been applied to AB Dor observations with SINFONI by Thatte et al. (2007). The latter show that the SNR of AB Dor C (a low-mass companion close to AB Dor A) can be increased by applying SD and using a priori information about the position of AB Dor C. They also give a general radial contrast profile based on the standard deviation of the final image. Translating their quoted $1 \sigma$ points into $3 \sigma$ and including them in Fig. \ref{sdi_img} implies that SD with SINFONI can reach almost as high contrast as NACO-SDI, if we generously assume that its SNR develops according to $t^{1/2}$. However, this contrast is only valid well outside of an angular separation quantified by Thatte et al. (2007) as the bifurcation radius. For the SINFONI ``H+K'' mode, the bifurcation radius is about 250 mas. Inside of this radius, a large fraction of the companion flux will inevitably be subtracted out by the SD technique (if its position is not known a priori or, equivalently, can be seen in the data already prior to applying the technique). Even outside of this radius, it is unclear whether the contrast is entirely applicable to blind searches, since a certain fraction of their flux would still be lost in the particular version of the SD technique upon which that result is based.

\begin{figure}[htb]
   \centering
   \includegraphics[width=8.5cm]{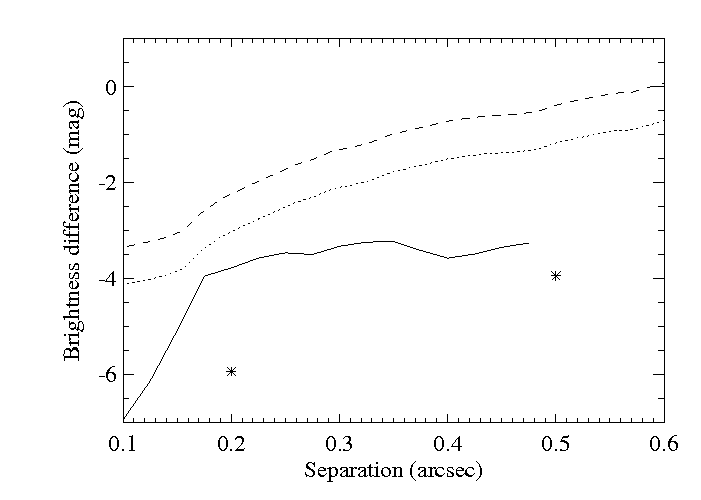}
\caption{Example comparison of physical and instrumental contrasts. The theoretical H-band contrast of a 2 $M_{\rm jup}$ companion to a 0.3 $M_{\rm sun}$ star at 100 Myr (see Baraffe et al. 1998 and Baraffe et al. 2003) has been translated into contrasts in the wavelength range of each respective instrument/method, and subtracted from the instrumental contrasts of Fig. \ref{sdi_err}. The planet would be detectable at any separation where an instrumental line is larger than 0 mag. NACO-SDI is clearly preferable for these cool companions (planet detectable at 0.6" and outwards). Solid line: SINFONI SDI. Dashed line: NACO-SDI. Dotted line: NACO-SDI at a common integration time. Stars: SINFONI SD.}
\label{sdi_phys_err}
    \end{figure}

Of importance for high-contrast imaging purposes is not only the instrumental contrast that can be achieved, but also the spectral energy distribution of the possible companion. Here, the SDI technique has a strong advantage over the SD technique for cool companions such as low-mass brown dwarfs and exoplanets. Such objects exhibit strongly increasing absorption in the H- and K-bands with decreasing temperature. According to, e.g., the spectral models of Burrows et al. (2003), for young planets, the flux is strongly concentrated directly shortwards of the methane feature at 1.6 $\mu$m in the H-band, and practically no flux is present in the K-band. This favors the SDI technique, since the off-methane filters are located precisely at the peak of the planetary flux. At the same time, it disfavors the SD technique, since it includes a large spectral range that practically contains only noise from the primary and no signal from the companion. To give a practical example of this, we use the Baraffe et al. (2003) and Burrows et al. (2003) theoretical models to model the actual physical contrast that the SDI and SD methods will face, respectively, in the case of a 2 $M_{\rm jup}$ companion to a 0.3 $M_{\sun}$ star at an age of 100 Myr. The same method is used in, e.g., Janson et al. (2007) and Apai et al. (2007). The physical contrast is subtracted from the instrumental (achieved contrast), and the results are shown in Fig. \ref{sdi_phys_err}. It is clear that SD is considerably less well applicable than SDI for such cool objects, at least in its present form. Obviously, this difference is less problematic for hotter objects. Still, high-contrast searches in general primarily aim to detect cool objects, and the SDI method is clearly preferable for such an objective. Since NACO-SDI performs better than SINFONI in this regard, this speaks in favor of the former for blind companion searches. In any case, for the SD technique to reach its full potential, it is necessary to know the position of the companion a priori. This underlines the fact that SINFONI is better suited for follow-up observations than for blind searches -- a purpose for which other instruments (e.g., NACO-SDI) are more appropriate.

As a further comment concerning the comparison between SINFONI MSDI and NACO-SDI for high-contrast imaging, we note that several practical issues speak against SINFONI in this regard. The bright object constraints are tighter for SINFONI than for NACO, and SINFONI does not provide an option to use neutral density filters; therefore, some of the most interesting objects for high-contrast imaging (such as $\epsilon$ Eri), which are observable with NACO, are not observable with SINFONI. The FOV is much smaller for SINFONI than for NACO-SDI at the same pixel scale (and even then, the pixel scale for SINFONI is 2 times coarser in one direction). Finally, since the data is more memory-demanding and complex than NACO-SDI data, it is also more cumbersome to reduce and work with.

The issues adressed here are of relevance for future generations of planet-finding instruments, since they indicate that in order to achieve high-quality results for SDI with an integral field unit, they have to be carefully designed for providing high AO performance, temporal stability, optical quality and minimized non-common path aberrations. Future high-contrast instruments, such as SPHERE and GPI, include units for integral field spectroscopy that are specifically designed for planet-finding purposes (as opposed to SINFONI, which is a general-purpose instrument), and should improve the capacity for SDI significantly. Meanwhile, it is generally easier to accommodate these requirements with a differential imager, hence plausibly, a better performance in this regard can be expected for a differential imager given the same cost and effort constraints. Hence, units such as these have an important role also for future instruments, possibly being advantageous for blind searches. In any case, given the capacity for attaining quantitative spectral information with integral field units, they are clearly preferable for follow-up searches, and are hence undeniably relevant for future instruments.

\subsection{Broad-band image}

\begin{figure*}[htb]
   \centering
   \includegraphics[width=16.0cm]{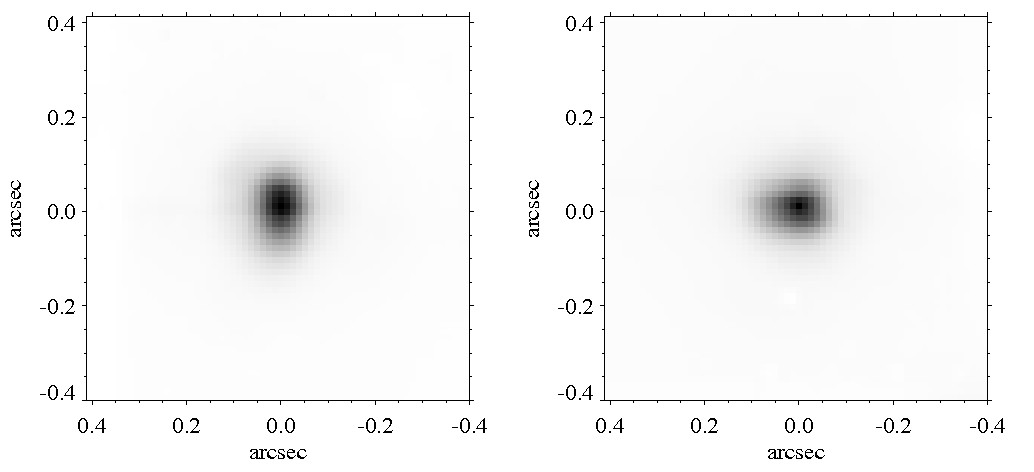}
\caption{Broad-band images covering the full wavelength range of L 449-1 at $0^{\circ}$ (left; North is up and East is to the left) and $90^{\circ}$ (right; North is to the right and East is up) rotation. While the images can be interpreted as suggesting the presence of a partially resolved companion, this is probably a PSF artefact.}
\label{coll_img}
    \end{figure*}

Two images collapsed over all wavelengths of the cubes of L 449-1 are shown in Fig. \ref{coll_img}, one at $0^{\circ}$ rotation and one at $90^{\circ}$. There is no convincing evidence for any low-contrast companion in those images. The PSF does look somewhat extended, and the extension rotates along with the camera, which is often indicative of a close binary. In this case, if such a secondary was real, it would be on the order of 10 \% of the brightness of the primary. However, it is likely to be a PSF artefact. A real physical companion that is fainter than the primary would inevitably be cooler. Hence, in a differential spectroscopy cube, where an average spectrum is subtracted from the original cube as described in the previous section, a feature would be seen at the expected position of the companion that is significantly redder than the rest of the field. In other words, it would be darker than average at short wavelengths, and brighter than average at long wavelengths. No such trend is seen in our differential spectroscopy cube, thus we conclude that the feature seen is probably an artefact. The effect can not be attributed to differential atmospheric refraction, as the position of the star does not systematically vary with wavelength in individual slices of the cube.

\subsection{Spectra}

The H- and K-band spectra of L 449-1 were divided by a standard star spectrum in order to remove telluric features. The standard star used for this purpose was HD 55397, which is classified as a B7 giant in SIMBAD. The spectrum of the standard star was divided by a blackbody spectrum corresponding to a temperature of 13000 K, as expected from its spectral type, in order to preserve the continuum of L449-1. HD 55397 has a few spectral features of its own, which would show up as apparent emission features in the final spectrum. To remove this effect, we replaced these spectral features in the standard star spectrum with interpolation of the surrounding continuum. Local telluric information is obviously lost in these particular places, which shows up as small deviations from standard spectra in the same spectral type range. 

\begin{figure*}[htb]
   \centering
   \includegraphics[width=14.0cm]{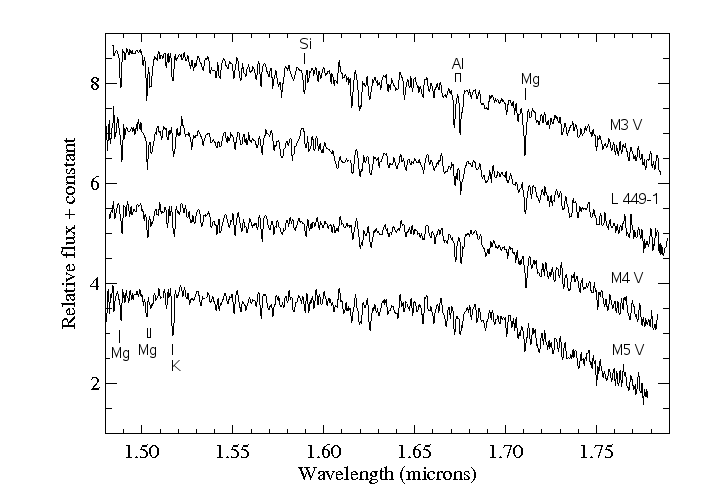}
\caption{Spectrum of L449-1 in the H-band, compared to M3, M4 and M5 standard spectra.}
\label{h_sptlock_all}
    \end{figure*}

\begin{figure*}[htb]
   \centering
   \includegraphics[width=14.0cm]{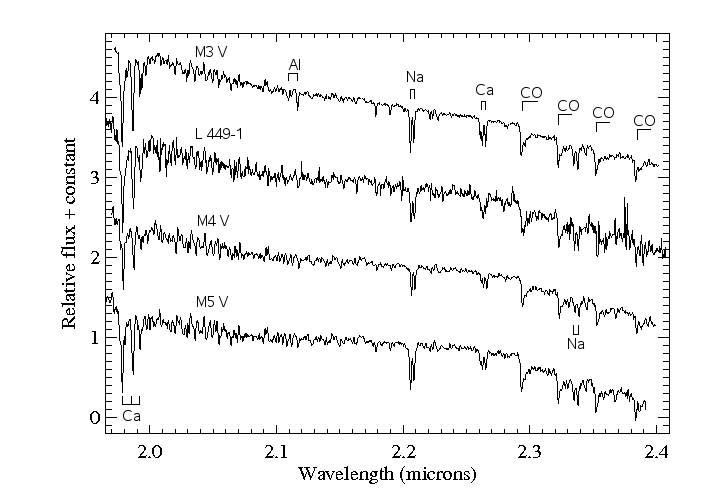}
\caption{Spectrum of L449-1 in the K-band, compared to M3, M4 and M5 standard spectra.}
\label{k_sptlock_all}
    \end{figure*}

We used the spectra to derive a spectral type, as a complement to the spectral analysis in the visual range by Scholz et al. (2005). For this purpose, we compared the H- and K-band spectra to the reference stars given in Cushing et al. (2005). The results are shown in Figs. \ref{h_sptlock_all} and \ref{k_sptlock_all}. Analysis of the most prominent absorption features places L449-1 in the range between M3 and M4 -- the Mg features near 1.5 $\mu$m, the K line at 1.54 $\mu$m, several FeH lines in the range of 1.6-1.7 $\mu$m, the Al doublet at 1.67-1.68 $\mu$m, the Mg line at 1.71 $\mu$m, the Ca features below 2.0 $\mu$m, the Al doublet at 2.11-2.12 $\mu$m, the Na doublet at 2.21 $\mu$m, the Ca feature at 2.27 $\mu$m and the CO bandheads at 2.3 $\mu$m and beyond are all consistently within this range. The spectral continuum is well preserved in integral field spectrographs, and also agrees well with a spectral type in the same range. Hence, we conclude that the spectral type is in the range of M3.0-M4.0, which agrees very well with the Scholz et al. (2005) characterization.

\section{Conclusions}

We have investigated the potential of the SDI technique with SINFONI for detecting cool substellar companions to stars. It was discovered that the performance is considerably worse than for NACO-SDI under very similar circumstances. In addition, the error is partly static, leading to a poor increase of performance with increasing integration time. The quasi-static nature of the noise, along with the fact that it is constant over most of the image space, implies that non-common path aberrations and errors from the slicing and reconstruction of the image dominate the residual noise. These results, along with practical considerations, clearly lead to the suggestion that ``blind'' differential imaging searches for cool companions (i.e., when no a priori information is available) are presently best performed with specialized instruments such as NACO-SDI, whereas follow-up observations, where more qualitative information is desired, are better suited for integral field units such as SINFONI. The results are highly relevant for the design of the next-generation instruments for planet detection.

While the collapsed broad-band images of L 449-1 imply the existence of a low-contrast candidate companion, the absence of a clear spectral signature associated with this feature suggests that it is a PSF artefact rather than a physical object.

The spectral type of L449-1, as determined from the H- and K-band spectra, was found to be in the range of M3.0-M4.0, in close agreement with previous results (M4, Scholz et al. 2005) in the visual range.

\begin{acknowledgements}
The authors wish to thank Alessandro Berton for his useful input during the observation preparations. M.J. gratefully receives financial support from IMPRS Heidelberg.
\end{acknowledgements}

\end{document}